\documentclass[12pt]{article}
\usepackage{amssymb, amsmath, amsfonts}

\newtheorem{theorem}{Theorem}
\newtheorem{lemma}[theorem]{Lemma}
\newtheorem{definition}[theorem]{Definition}

\newtheorem{example}[theorem]{Example}
\newtheorem{remark}[theorem]{Remark}

\begin{document}
\title{ On existence of an x-integral for a semi-discrete chain of hyperbolic type}

\author{  Kostyantyn Zheltukhin$^1$ and Natalya Zheltukhina$^2$\\
\small $^1$ Department of Mathematics, Faculty of Science,\\
\small Middle East Technical University 06531 Ankara, Turkey.\\
\small email:zheltukh@metu.edu.tr\\
\small $^2$ Department of Mathematics, Faculty of Science,\\
\small  Bilkent University, 06800 Ankara, Turkey.\\
\small email:natalya@fen.bilkent.edu.tr}

\begin{titlepage} 

\maketitle

\begin{abstract}
A class of semi-discrete chains of the form $t_{1x} = f(x, t, t_1, t_x)$ is considered. For the given
chains easily verifiable conditions for existence of x-integral of minimal order 4 are obtained.
\end{abstract}

\end{titlepage}

\section{Introduction}

In the present paper we consider the integrable differential-difference chains of hyperbolic type
\begin {equation} \label{basic_eqn}
t_{1x}=f(x,t,t_1,t_x),
\end{equation}
where the function $t(n,x)$ depends on discrete variable $n$ and continuous variable $x$. We use the following notations
$t_x=\frac{\partial }{\partial x} t$ and $t_1=t(n+1,x)$. It is also convenient to denote  $t_{[k]}=\frac{\partial ^k }{\partial x^k} t$,  $k\in {\mathbb N}$
and $t_m=t(n+m,x)$, $m\in {\mathbb Z}$.

The integrability of the chain (\ref{basic_eqn}) is understood as Darboux integrability that is existence of so called $x$- and $n$-integrals \cite{Dar, HP}. 
Let us give the necessary definitions.
\begin{definition}
 Function $F(x,t,t_1,\dots ,t_k)$ is called an $x$-integral of the equation (\ref{basic_eqn}) if
$$
 D_x F(x,t,t_1,\dots ,t_k)=0
 $$
for all solutions of (\ref{basic_eqn}). The operator $D_x$ is the total derivative with respect to $x$.
\end{definition}

\begin{definition}
 Function $G(x,t,t_x,\dots ,t_{[m]})$ is called an $n$-integral of the equation (\ref{basic_eqn}) if
 $$
 D G(x,t,t_x,\dots ,t_{[m]})=G(x,t,t_x,\dots ,t_{[m]})
 $$
for all solutions of (\ref{basic_eqn}). The operator $D$ is a shift operator.
\end{definition}
To show the existence of $x$- and $n$-integrals we can use the notion of characteristic ring. The notion of characteristic ring was  introduced by Shabat to study   hyperbolic systems of exponential type (see \cite{ShYa}). This approach turns out to be very convenient to study and classify the integrable equations of hyperbolic type (see \cite{ZhMHSh} and references there in).

For difference and differential-difference chains the notion of characteristic ring was developed by Habibullin (see \cite{H}-\cite{HZhP3}). In particular, in \cite{HP} the following theorem was proved

\begin{theorem} (see \cite{HP}).
 A chain (\ref{basic_eqn}) admits a non-trivial $x$-integral if and only if its
characteristic x-ring is of finite dimension.\\
 A chain (\ref{basic_eqn}) admits a non-trivial n-integral if and only if its
characteristic $n$-ring is of finite dimension.
\end{theorem}
For known examples of integrable chains the dimension of the characteristic ring is small. The differential-difference chains with three dimensional characteristic $x$-ring were considered in 
\cite{HZhP1}. We consider  chains with four dimensional characteristic $x$-ring, such chains admit $x$-integral of minimal order four. That is we obtain  necessary and sufficient conditions for a chain to have a four dimensional characteristic $x$-ring. This conditions can be easily checked by direct calculations.

Note that if a chain (\ref{basic_eqn})
admits a nontrivial $x$-integral $F(x,t,t_1,\dots t_k)$ and a non trivial $n$-integral $G(x,t,t_x,\dots ,t_{[m]})$ its solutions satisfy two ordinary equations
$$
F(x,t,t_1,\dots ,t_k)=a(n),
$$
$$
 G(x,t,t_x,\dots ,t_{[m]})=b(x)
$$
for some functions $a(n)$ and $b(x)$. This allows to solve (\ref{basic_eqn}) (see \cite{HZhS}).

The paper is organized as follows. In Section 2 we derive necessary and sufficient conditions on function $f(x,t,t_1,t_x)$ so that the chain (\ref{basic_eqn}) has four dimensional characteristic ring
and in Section 3 we consider some applications of the derived conditions.  

\section{Chains admitting four dimensional $x$-algebra.}

Suppose  $F$ is an $x$-integral of the chain (\ref{basic_eqn}) then its positive shifts  and negative shifts $D^k\, F$, $k\in{\mathbb Z}$,  are also $x$-integrals. So, looking for an   $x$-integral it is convenient to assume that it depends on positive and negative shits of $t$.

To express $x$ derivatives of negative shifts we can apply $D^{-1}$ to the chain (\ref{basic_eqn}) and obtain
$$
t_x=f(x,t_{-1},t,t_x).
$$
Solving the above  equation for $t_{-1x}$ we get
$$
t_{-1x}=g(x,t_{-1},t,t_x).
$$

Let $F(x,t,t_1, t_{-1},\dots)$ be an $x$-integral of the chain (\ref{basic_eqn}).
Then on solutions of (\ref{basic_eqn}) we have
$$
 D_x F= \frac{\partial F}{\partial x} + t_x\frac{\partial F}{\partial t}+ t_{1x}\frac{\partial F}{\partial t_1}+ t_{-1x}\frac{\partial F}{\partial t_{-1}}+t_{2x}\frac{\partial F}{\partial t_2}+ t_{-2x}\frac{\partial F}{\partial t_{-2}}+\dots=0
$$
or 
$$
 D_x F= \frac{\partial F}{\partial x} + t_x\frac{\partial F}{\partial t}+ f\frac{\partial F}{\partial t_1}+ g\frac{\partial F}{\partial t_{-1}}+Df\frac{\partial F}{\partial t_2}+ D^{-1}g\frac{\partial F}{\partial t_{-2}}+\dots=0.
$$

Define a vector field
 \begin{equation}
 K=\frac{\partial}{\partial x} + t_x\frac{\partial}{\partial t}+ f\frac{\partial}{\partial t_1}+ g\frac{\partial}{\partial t_{-1}}+Df\frac{\partial}{\partial t_2}+D^{-1} g\frac{\partial}{\partial t_{-2}}+\dots,
\end{equation}
then 
$$
 D_x F=K \,F.
$$
Note that $F$  does not depend on $t_x$ but the coefficients of $K$ do depend on $t_x$.  So we introduce a vector field
 \begin{equation}
 X=\frac{\partial}{\partial t_{x}}
\end{equation}
The vector fields $K$ and $X$ generate the characteristic $x$-ring~$L_x$.

Let us  introduce  some other vector fields from $L_x$.
\begin{equation}
C_1=[X,K] \quad \mbox{and} \quad  C_n=[X,C_{n-1}] \quad n=2,3,\dots
\end{equation}
and
\begin{equation}
Z_1=[K,C_1] \quad \mbox{and} \quad  Z_n=[K,Z_{n-1}] \quad n=2,3,\dots
\end{equation}
Thus
$$
\begin{array}{l}
 \displaystyle{C_1=\frac{\partial}{\partial t} + f_{t_x}\frac{\partial}{\partial t_1}+ g_{t_x}\frac{\partial}{\partial t_{-1}}+\dots}\\
 \\
 \displaystyle{C_2= f_{t_xt_x}\frac{\partial}{\partial t_1}+ g_{t_xt_x}\frac{\partial}{\partial t_{-1}}+\dots}\\
\\
 \displaystyle{Z_1=\left(f_{t_{xx}}+t_xf_{t_xt}+ff_{t_xt_1}-f_t-f_{t_x}f_{t_1}\right) \frac{\partial}{\partial t_1}+ \left(g_{t_{xx}}+t_xg_{t_xt}+gg_{t_xt_1}-g_t-g_{t_x}g_{t_1}\right)\frac{\partial}{\partial t_{-1}}+\dots}\\
\end{array}
$$
and so on.

It is easy to see that if  $f_{t_xt_x}\ne 0$ then the vector fields $X$, $K$, $C_1$ and $C_2$ are  linearly independent and must form a basis of  $L_x$ provided $dim L_x=4$. 
By Lemma 3.6 in \cite {HZhP1}, if $f_{t_x t_x}=0$ and $\displaystyle{\left(f_{t_{xx}}+t_xf_{t_xt}+ff_{t_xt_1}-f_t-f_{t_x}f_{t_1}\right)}=0$ then  $dim L_x=3$. So in the case $f_{t_xt_x}=0$ we may assume 
$\displaystyle{\left(f_{t_{xx}}+t_xf_{t_xt}+ff_{t_xt_1}-f_t-f_{t_x}f_{t_1}\right)}\ne 0$. Then the vector fields $X$, $K$, $C_1$ and $Z_1$ are  linearly independent and must form a basis of  $L_x$ provided $dim L_x=4$. We consider this two cases separately.

In the rest of the paper we assume that the characteristic ring $L_x$ is four dimensional.

\begin{remark}\label{rem1}
It is convenient to check equalities between vector fields using the automorphism $D(\,\,)D^{-1}$. Direct calculations show that
$$
DXD^{-1}=\frac{1}{f_x}X,
$$
$$
DKD^{-1}=K-\frac{f_x+t_xf_t+ff_{t_1}}{f_{t_x}} X.
$$
The images of other vector fields under this automorphism can be obtained by commuting  $DXD^{-1}$ and $DKD^{-1}$.
\end{remark}

\subsection{$f(x,t,t_1, t_x)$ is non linear with respect to $t_x$.}

Let $f(x,t,t_1, t_x)$ be non linear with respect to $t_x$, $f_{t_xt_x}\ne 0$. Then the vector fields $X$, $K$, $C_1$ and $C_2$ form a basis of $L_x$.
For the algebra $L_x$ to be spanned by  $X$, $K$, $C_1$ and $C_2$ it is enough that $C_3$ and $Z_1$ are linear combinations of $X$, $K$, $C_1$ and $C_2$.
From the form of the vector fields  it follows that we must have
$$
 C_3=\lambda C_2 \quad \mbox{and} \quad  Z_1=\mu C_2
$$
for some functions $\mu$ and $\lambda$. The conditions for the above equalities to hold are given by the following theorem.  

\begin{theorem}\label{th1}
  The chain (\ref{basic_eqn}) with $f_{t_xt_x}\ne 0$ has characteristic ring $L_x$ of dimension four if and only if the following conditions hold
\begin{equation}\label{L_4first_cond}
D \left(\frac{f_{t_xt_xt_x}}{f_{t_xt_x}}\right)=\frac{f_{t_xt_xt_x}f_{t_x}-3f^2_{t_xt_x}}{f_{t_xt_x}f^2_{t_x}}.
\end{equation}
\begin{equation}\label{L_4second_cond}
\begin{array}{l}
 \displaystyle{D \left(\frac{f_{xt_x}+t_xf_{t_xt}+ff_{t_xt_1}-f_{t}-f_{t_x}f_{t_1}}{f_{t_xt_x}}\right)=}\\
 \\
\displaystyle{\hspace{2cm}\frac{f_{xt_x}+t_xf_{t_xt}+ff_{t_xt_1}-f_t-f_{t_x}f_{t_1}}{f_{t_xt_x}}f_{t_x}-(f_x+t_xf_t+f_{t_1}).}
\end{array}
\end{equation}
The characteristic ring is generated by the vector fields $X,K,C_1,C_2$.
\end{theorem}
{\bf Proof.} By Remark \ref{rem1} we have
$$
DC_2D^{-1}=\frac{1}{f^2_{t_x}}C_2-\frac{f_{t_xt_x}}{f^3_{t_x}}C_1+\frac{f_{t_xt_x}f_{t}}{f^4_{t_x}} X
$$
$$
DC_3D^{-1}=\frac{1}{f^3_{t_x}} C_2 -\frac{3f_{t_xt_x}}{f^4_{t_x}}C_2-\frac{f_{t_xt_xt_x}f_{t_x}-3f^2_{t_xt_x}}{f^5_{t_x}}C_1+f_t\frac{f_{t_xt_xt_x}f_{t_x}-3f^2_{t_xt_x}}{f^6_{t_x}}X
$$
$$
DZ_1D^{-1}=\frac{1}{f_{t_x}}Z_1-\left(\frac{mf_{t_x}+p}{f^2_{t_x}}\right)\left( C_1-\frac{f_t}{f_{t_x}} X \right),
$$
where $\displaystyle{p=\frac{f_x+t_xf_t+ff_{t_1}}{f_{t_x}}}$ and $m=\displaystyle{\frac{-(f_{xt_x}+t_xf_{t_xt}+ff_{t_xt_1})+f_t+f_{t_x}f_{t_1}}{f_{t_x}}}$.

\noindent
The equality $ C_3=\lambda C_2 $ implies that  
\begin{equation}\label{eq1}
DC_3D^{-1}=(D\lambda) \, DC_2D^{-1}.
\end{equation}
Substituting expressions for $DC_2D^{-1}$ and $DC_3D^{-1}$ into (\ref{eq1}) and comparing coefficients of $C_1$, $C_2$ and $X$ we obtain that $\lambda$ satisfies
$$
\lambda=f_{t_x}(D\lambda) +\frac{3f_{t_xt_x}}{f_{t_x}}
$$
$$
(D\lambda)=\frac{f_{t_xt_xt_x}f_{t_x}-3f^2_{t_xt_x}}{f_{t_xt_x}f^2_{t_x}}.
$$
We can find $\lambda$ and $D\lambda$ independently and condition that $D\lambda$ is a shift of $\lambda$ leads to (\ref{L_4first_cond}). 

\noindent
The equality $Z_1=\mu C_2$ implies that  
\begin{equation}\label{eq2}
DZ_1D^{-1}=(D\mu) \, DC_2D^{-1}.
\end{equation}
Substituting expressions for $DC_2D^{-1}$ and $DC_3D^{-1}$ into (\ref{eq2}) and comparing coefficients of $C_1$, $C_2$ and $X$ we obtain that $\mu$ satisfies
$$
\mu-\frac{f_x+t_xf_{t}+ff_{t_1}}{f_{t_x}}=\frac{(D \mu)}{f_{t_x}}
$$
and
$$
-(f_{xt_x}+t_xf_{t_xt}+ff_{t_xt_1}-f_t-f_{t_x}f_{t_1}) +\frac{f_x+t_xf_t+ff_{t_1}}{f_{t_x}}f_{t_xt_x}=-\frac{f_{t_xt_x}(D\mu)}{f_{t_x}}
$$
We can find $\mu$ and $D\mu$ independently and condition that $D\mu$ is a shift of $\mu$ leads to (\ref{L_4second_cond}). $\Box$

\begin{remark}
Let $dim \,L_x=4$ and $f_{t_{xx}}\ne 0$. Then the characteristic ring $L_x$  have the following multiplication table  
\begin{center}
\begin{tabular}{r|rrrr}
 & $X$ & $K$ & $C_1$ & $C_2$ \\
 \hline
 $X$ & $0$ & $C_1$ & $C_2$ & $\mu C_2$ \\
 $K$ & $-C_1$& $0$ & $\lambda C_2$ & $\rho C_2$ \\
$C_11$ & $-C_2$ & $-\lambda C_2$ & $0$ & $\eta C_2$  \\
$C_2$ & $-\mu C_2$ & $-\rho C_2$ & $-\eta C_2$ & $0$ \\
\end{tabular}
\end{center}
where  $\rho=\lambda\mu +X(\lambda)$ and $\eta=X(\rho)-K(\mu)$.
\end{remark}

\begin{example}
Consider the following chain 
$$
t_{1x}=\frac{tt_x-\sqrt{t^2_x-M^2}(t_1+t)}{t_1}
$$
introduced by Habibullin and  Zheltukhina \cite{HZh}.
We can easily check that the function 
$$
f(t,t_1,t_x)=\frac{tt_x-\sqrt{t^2_x-M^2}(t_1+t)}{t_1}
$$
satisfies the conditions of Theorem \ref{th1}. Hence the corresponding $x$-algebra is four dimensional.
The chain has the following $x$-integral 
$$
F=\frac{(t_1^2-t^2)(t_1^2-t_2^2)}{t_1^2}.
$$
\end{example}

\subsection{$f(x,t,t_1, t_x)$ is  linear with respect to $t_x$.}

Let $f(x,t,t_1, t_x)$ be linear with respect to $t_x$, $f_{t_xt_x}= 0$. 
Then vector fields $X$, $K$, $C_1$ and $Z_1$  form a basis of $L_x$. The condition $f_{t_xt_x}= 0$ also  implies that the vector field $C_2=0$, see \cite{HZhP1}. 
For the algebra $L_x$ to be spanned by  $X$, $K$, $C_1$ and $Z$ it is enough that  $Z_2$ is a linear combination of $X$, $K$, $C_1$ and $Z_1$.
From the form of the vector fields  it follows that we must have
$$
 Z_2= \alpha Z_1
$$
for some function $\alpha$. The conditions for the above equality to hold  given by the following theorem.  

\begin{theorem}\label{th2}
 The chain (\ref{basic_eqn}) with $f_{t_xt_x}= 0$ has the characteristic ring $L_x$ of dimension four if and only if the following condition hold
\begin{equation}\label{L_4_cond_case_2}
D \left(\frac{K(m)}{m}-m+\frac{f_t}{f_{t_x}}\right)=\frac{K(m)}{m}+m-f_{t_1}.
\end{equation}
where $m=\displaystyle{\frac{-(f_{xt_x}+t_xf_{t_xt}+ff_{t_xt_1})+f_t+f_{t_x}f_{t_1}}{f_{t_x}}}$.
The characteristic ring is generated by the vector fields $X,K,C_1,Z_1$.
\end{theorem}
{\bf Proof.} By Remark \ref{rem1} we have
$$
DZ_1D^{-1}=\frac{1}{f_{t_x}}Z_1-\left(\frac{mf_{t_x}+p}{f^2_{t_x}}\right)\left( C_1-\frac{f_t}{f_{t_x}} X \right),
$$
and
$$
DZ_2D^{-1}=\left(K\left(\frac{1}{f_{t_x}}\right)+\frac{\alpha+m}{f_{t_x}}\right)Z_1+\left(K\left(\frac{m}{f_{t_x}}\right) + \frac{mf_t}{f^2_{t_x}}-pX\left(\frac{m}{f_{t_x}}\right)\right)\left(C_1-\frac{f_t}{f_{t_x}} X  \right) 
$$
The equality $ Z_2=\alpha Z_1 $ implies that  
\begin{equation}\label{eq2.1}
DZ_2D^{-1}=(D\alpha) \, DZ_1D^{-1}.
\end{equation}
Substituting expressions for $DZ_1D^{-1}$ and $DZ_2D^{-1}$ into (\ref{eq2.1}) and comparing coefficients of $C_1$, $Z_1$ and $X$ we obtain that $\alpha$ and $D(\alpha)$ satisfy
$$
K\left(\frac{1}{f_{t_x}}\right)+\frac{m}{f_{t_x}}+\frac{\alpha}{f_{t_x}}=\frac{D(\alpha)}{f_{t_x}}
$$
$$
K\left(\frac{m}{f_{t_x}}\right)+\frac{mf_t}{f^2_{t_x}}=\frac{mD(\alpha)}{f_{t_x}}
$$
We can find $\alpha$ and $D(\alpha)$ independently and condition that $D(\alpha)$ is a shift of $\alpha$ leads to (\ref{L_4_cond_case_2}). $\Box$

\begin{remark}
Let $dim \,L_x=4$ and $f_{t_{xx}}=0$. Then the characteristic ring $L_x$  have the following multiplication table  
\begin{center}
\begin{tabular}{r|rrrr}
 & $X$ & $K$ & $C_1$ & $Z_1$ \\
 \hline
 $X$ & $0$ & $C_1$ & $0$ & $0$ \\
 $K$ & $-C_1$& $0$ & $Z_1$ & $\alpha Z_1$ \\
$C1$ & $0$ & $-Z_1$ & $0$ & $X(\alpha)Z_1$  \\
$Z_1$ & $0$ & $-\alpha Z_1$ & $-X(\alpha) Z_1$ & $0$ \\
\end{tabular}
\end{center}
\end{remark}

\begin{example}
Consider the following chain 
$$
t_{1x}=t_x +e^{\frac{t+t_1}{2}}
$$
introduced by Adler and Startsev \cite{AS}.
We can easily check that the function 
$$
f(t,t_1,t_x)=t_x +e^{\frac{t+t_1}{2}}
$$
satisfies the conditions of Theorem \ref{th2}. Hence the corresponding $x$-algebra is four dimensional.
The chain has the following $x$-integral 
$$
F=e^\frac{t_1-t}{2}+e^\frac{t_1-t_2}{2}
$$
\end{example}

\section{Applications}

The conditions derived in the previous section  can be used to determine some restrictions on the form of the function $f(x,t,t_1,t_x)$ in (\ref{basic_eqn}).
 
 \begin{lemma}
 Let the chain (\ref{basic_eqn}) have four dimensional characteristic $x$-ring. Then
 \begin{equation}\label{form_for_f1}
 f=M(x,t,t_x)A(x,t,t_1)+t_xB(x,t,t_1) +C(x,t,t_1),
 \end{equation}
 where $M,\,A,\,B$ and $C$ are some functions.
 \end{lemma}

{\bf Proof.}
Let $f_{t_xt_x}\ne 0$ (if $f_{t_xt_x}=0$ then $f$ obviously has the above form). Since  characteristic $x$-ring has dimension four the condition (\ref{L_4first_cond}) holds. 
It is easy to see that  (\ref{L_4first_cond}) implies that $\displaystyle{\frac{f_{t_xt_xt_x}}{f_{t_xt_x}}}$ does not depend on $t_1$. Hence 
$$
X(\ln |f_{t_xt_x}|)=M_1(x,t,t_x)\quad \mbox{and}\quad \ln |f_{t_xt_x}|=M_2(x,t,t_x)+A_1(x,t,t_1).
$$
The last equality implies (\ref{form_for_f1}). $\Box$

We can also put some restrictions on the shifts of the function $f(x,t,t_1,t_x)$ in (\ref{basic_eqn}).
  \begin{lemma}
 Let the chain (\ref{basic_eqn}) have four dimensional characteristic $x$-ring and $f_{t_xt_x}\ne 0.$ Then 
 \begin{equation}\label{form_for_f2}
 Df=-H_1(x,t,t_1,t_2)t_x+H_2(x,t,t_1,t_2)f+H_3(x,t,t_1,t_2),
 \end{equation}
  where $H_1,\,H_2$ and $H_3$ are some functions.
 \end{lemma}

{\bf Proof.}
Note that the shift operator $D$ and the vector field $X$ satisfy
\begin{equation}\label{D}
DX=\frac{1}{f_{t_x}}XD.
\end{equation}
The condition (\ref{L_4first_cond}) can be written as
$$
DX(\ln |f_{t_xt_x}|)=\frac{1}{f_{t_x}} X(\ln |f_{t_xt_x}|-\ln |f_{t_x}|^3)
$$
Using (\ref{D}) we get
$$
\frac{1}{f_{t_x}}XD(\ln |f_{t_xt_x}|)=\frac{1}{f_{t_x}} X(\ln |f_{t_xt_x}|-\ln |f_{t_x}|^3)
$$
which implies that
$$
X\left( \ln \left| f_{t_x}^3\frac{Df_{t_xt_x}}{f_{t_xt_x}}\right|\right)=0\quad \mbox{or}\quad X\left(  f_{t_x}^3\frac{Df_{t_xt_x}}{f_{t_xt_x}}\right)=0.
$$
Thus $\displaystyle{Df_{t_xt_x}=H_1(x,t,t_1,t_2)\frac{f_{t_xt_x}}{f^3_{t_x}}}$.
Since $Df_{t_xt_x}=DX(f_{t_x})$ and $\frac{f_{t_xt_x}}{f^3_{t_x}}=-\frac{1}{f_{t_x}}X(\frac{1}{f{t_x}})$ we can rewrite previous equality using (\ref{D}) as
$$
X\left(Df_{t_x}+H_1(x,t,t_1,t_2)\frac{1}{f_{t_x}}\right)=0
$$
which implies
$$
Df_{t_x}=-H_1(x,t,t_1,t_2)\frac{1}{f_{t_x}}+H_2(x,t,t_1,t_2).
$$
Writing 
$$
DX(f)=-H_1(x,t,t_1,t_2)\frac{1}{f_{t_x}}+H_2(x,t,t_1,t_2)\frac{f_{t_x}}{f_{t_x}}
$$
and applying (\ref{D}) as before we get
$$
X(Df+H_1(x,t,t_1,t_2)t_x-H_2(x,t,t_1,t_2)f)=0.
$$
The last equality gives (\ref{form_for_f2}). $\Box$

Note that the equality (\ref{form_for_f2}) can be written as 
$$
t_{2x}=H_2(x,t,t_1,t_2)t_{1x}-H_1(x,t,t_1,t_2)t_x+H_3(x,t,t_1,t_2).
$$

\end{document}